\documentclass[%
 reprint,
 amsmath,amssymb,superscriptaddress,
 prb,
]{revtex4-1}

\usepackage{graphicx}
\usepackage{dcolumn}
\usepackage{bm}
\usepackage{color}
 
\begin{document}


\title{Polarization-entangled twin photons from two-photon quantum-dot emission}

\author{Dirk Heinze}
\affiliation{Department of Physics and Center for Optoelectronics and Photonics Paderborn (CeOPP), University of Paderborn, Warburger Strasse 100, 33098 Paderborn, Germany. }%
\author{Artur Zrenner}
\affiliation{Department of Physics and Center for Optoelectronics and Photonics Paderborn (CeOPP), University of Paderborn, Warburger Strasse 100, 33098 Paderborn, Germany. }%
\author{Stefan Schumacher}%
\affiliation{Department of Physics and Center for Optoelectronics and Photonics Paderborn (CeOPP), University of Paderborn, Warburger Strasse 100, 33098 Paderborn, Germany. }%
\affiliation{College of Optical Sciences, University of Arizona, Tucson, Arizona 85721, USA.}

\date{\today}

\begin{abstract}
Semiconductor quantum dots are promising sources for polarization-entangled photons. As an alternative to the usual cascaded biexciton-exciton emission, direct two-photon emission from the biexciton can be used. With a high-quality optical resonator tuned to half the biexciton energy, a large proportion of the photons can be steered into the two-photon emission channel. In this case the degree of polarization entanglement is inherently insensitive to the exciton fine-structure splitting. In the present work we analyze the biexciton emission with particular emphasis on the influence of coupling of the quantum-dot cavity system to its environment. Especially for a high-quality cavity, the coupling to the surrounding semiconductor material can open up additional phonon-assisted decay channels. Our analysis demonstrates that with the cavity tuned to half the biexciton energy, the potentially detrimental influence of the phonons on the polarization entanglement is strongly suppressed -- high degrees of entanglement can still be achieved. We further discuss spectral properties and statistics of the emitted twin photons.  

\end{abstract}

\pacs{Valid PACS appear here}
\maketitle


\section{Introduction}

Photon pairs with a high degree of polarization entanglement are a key ingredient to a number of quantum information protocols.\cite{RevModPhys.74.145,Knill2001,Bouwmeester1997,Gisin2007} Semiconductor quantum dots have proven their capabilities for on-demand generation of individual pairs of such polarization-entangled photons.\cite{Muller2014} However, in most high-quality semiconductor quantum-dot structures, the fine-structure splitting between exciton levels limits the achievable degree of polarization entanglement.\cite{PhysRevB.74.235310} Recent achievements show that this obstacle can be overcome by applying strain to the quantum dot \cite{PhysRevLett.114.150502,Zhang2015, Chen2016} or by selecting those quantum dots on a given sample that possess a particularly small fine-structure splitting.   
As an alternative approach, it was shown theoretically that a high degree of polarization-entanglement can also be obtained by using a direct two-photon emission process from the quantum-dot biexciton inside an optical cavity -- independent of the fine-structure splitting.\cite{Schumacher2012}  
 
In quantum-dot-cavity systems where strong coupling between the electronic transitions and the optical cavity mode is realized,  two photon transitions gain importance.\cite{Heinze2015,delValleBiexDressing} \textcolor{black}{Starting with the proposal  \cite{1367-2630-13-11-113014} and experimental demonstration \cite{PhysRevLett.107.233602} of cavity enhanced two-photon emission, first potential applications have been analyzed \cite{triggerdTwinPhotonSource,Kuhn:16}.} Also interaction with phonons is particularly important in cavity systems.\cite{PhysRevLett.110.147401,PhysRevB.85.121302,PhysRevLett.114.137401,PhysRevB.90.241404,PhysRevB.91.161302} Phonon mediated processes have been shown to contribute to an asymmetric spectral line-broadening\cite{PhysRevB.85.115309} and phonon-assisted cavity feeding\cite{PhysRevB.81.155303}. The precise influence of phonon-mediated processes depends on the cavity detuning and phonon-bath properties.\cite{PhysRevX.1.021009} Also the two-photon emission from the biexciton may be influenced by this mechanism. Here, the question arises if longitudinal acoustic (LA) phonons can affect the system dynamics in such a way that the degree of entanglement of twin photons emitted from the biexciton is significantly reduced.  

\begin{figure}[t] 
\centering
\includegraphics[scale=0.25]{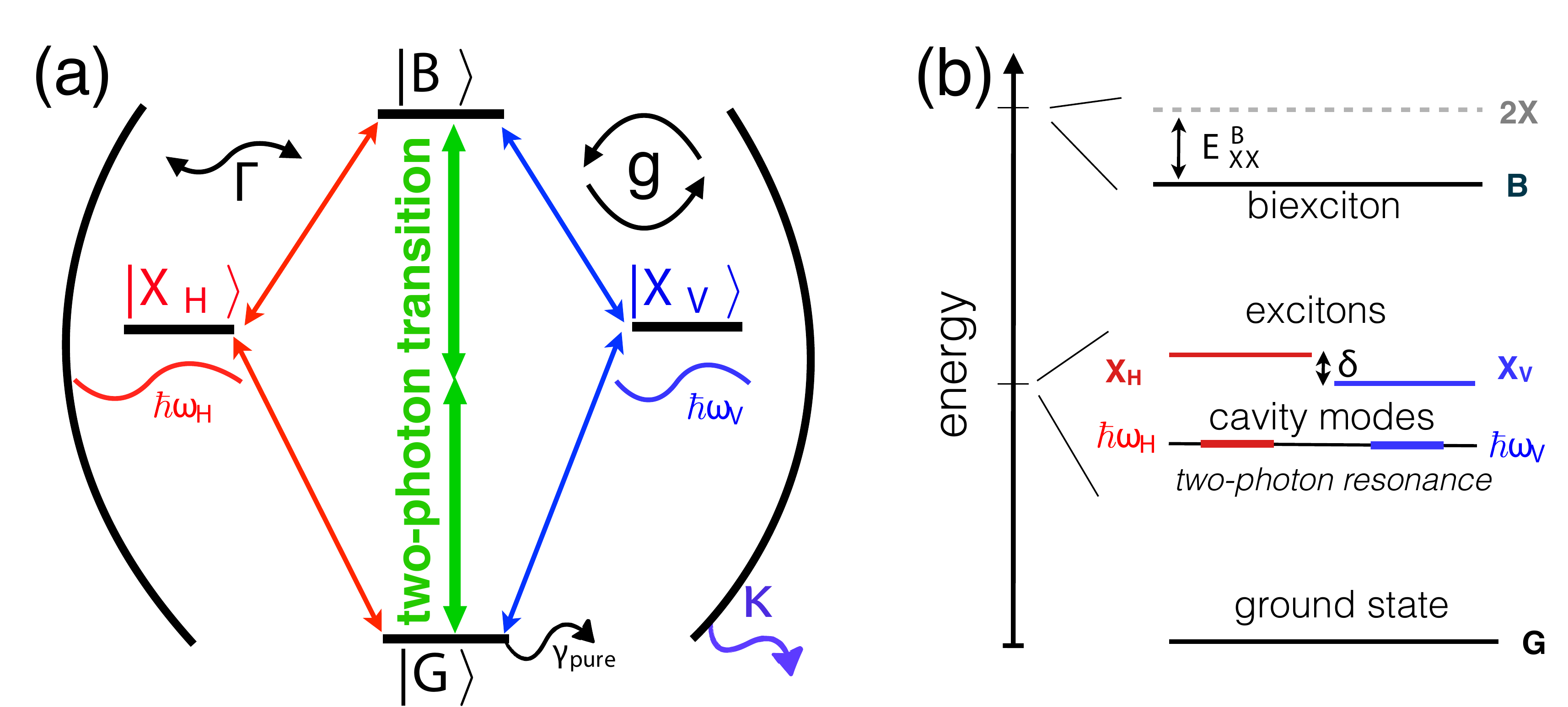}
\caption{Schematics of the quantum-dot cavity system.  a) The electronic levels are shown for the biexciton ($B$) and the two excitons ($X_V$, $X_H$) and the corresponding cavity modes with energies ($\hbar \omega_V$,$\hbar \omega_H$) and the ground state ($G$). The excitonic coherences decay with a  pure dephasing rate of $\gamma_{\mathrm{pure}}$ and the photons leave the cavity with $\kappa$, respectively. Phonon-assisted processes are denoted by $\Gamma$. b) Energy scheme for the system in a). The biexciton binding-energy $E_{XX}^B$ is typically of the order of meV and the electronic fine-structure splitting $\delta$ is typically of the order of several $\mathrm{\mu eV}$. The cavity modes are tuned  \textcolor{black}{to} the resonance of the degenerate two-photon transition  \textcolor{black}{ to provide for twin photon generation}.}
\label{QD_Cav}
\end{figure}

In this paper we analyze the effect of phonon mediated cavity (bi)exciton coupling and present a detailed analysis based on a Born-Markov approximation\cite{PhysRevB.65.235311,PhysRevB.85.115309} for the coupling to the phonon bath. In the emission scheme studied, the cavity is tuned near half the biexciton energy. In this case phonon-assisted cavity feeding primarily occurs through phonon absorption, which is significantly suppressed at low temperatures. Our results show that even including phonon-assisted processes, high degrees of polarization entanglement can still be achieved from a direct two-photon emission from the biexciton. We also analyze spectral properties and statistics of the emitted photons.

\section{Theory}
\label{Ref-Theory}
 
In this section an introduction to the theoretical description of the quantum-dot cavity system is given. This includes the general formulation of the theory, the formulation of a master equation describing the system dynamics, the system's coupling to the environment including phonon-assisted processes, and a short discussion of the two-photon density matrix needed to study the quantum properties of the emitted photons. 

\subsection{The Quantum-Dot Cavity System}
\label{System}

The present work is focussed on the photon emission from the biexciton in a semiconductor quantum dot inside an optical resonator. To analyze this process theoretically, we model the quantum-dot cavity system as a system of four electronic configurations and two orthogonal photon cavity modes. The states of the system are schematically depicted in Fig.~\ref{QD_Cav}. The electronic configurations included are the biexciton (B), the two excitons ($\mathrm{X_V}$, $\mathrm{X_H}$), and the ground state (G). The two orthogonal cavity modes are at energies $\hbar \omega_V$ and $\hbar \omega_H$. The biexciton binding energy is $E^B_{XX}$ and the exciton levels are split by a fine-structure splitting $\delta$. The Hamiltonian of this system is given by:\cite{PhysRevB.81.195319}
\begin{equation}
\begin{split}
H_S=E_G \vert G \rangle \langle G \vert + E_H \vert X_H \rangle \langle X_H \vert + E_V \vert X_V \rangle \langle X_V \vert  \\ + E_B \vert B \rangle \langle B \vert 
+ \sum_{i=H,V}  \hbar \omega_i b_i^{\dagger} b_i  + [\mathcal{X} + h.c. ]\,. 
\end{split}
\label{Hamiltonian}
\end{equation}
It contains the free energies of the electronic states of the quantum dot and the photons inside the cavity. The photon creation and annihilation operators are denoted by $b_i^{\dagger}$ and $b_i$, with $i=H,V$, respectively. The interaction part with coupling constant $g$  \textcolor{black}{ is given by}
\begin{eqnarray}
\begin{split}
\mathcal{X} = &-g[( \vert X_V \rangle \langle G \vert b_V - \vert B \rangle \langle X_V \vert b_V ) ]& \\ 
&-g[( \vert X_H \rangle \langle G \vert b_H + \vert B \rangle \langle X_H \vert b_H )  ].&  
\end{split}
\label{InteractionQDCav}
\end{eqnarray}
 \textcolor{black}{and} represents the coupling of the cavity-modes and the electronic system and induces transitions between electronic state while emitting or absorbing photons.

\subsection{The Master Equation}

\begin{figure}[t]
\centering
\includegraphics[scale=0.3]{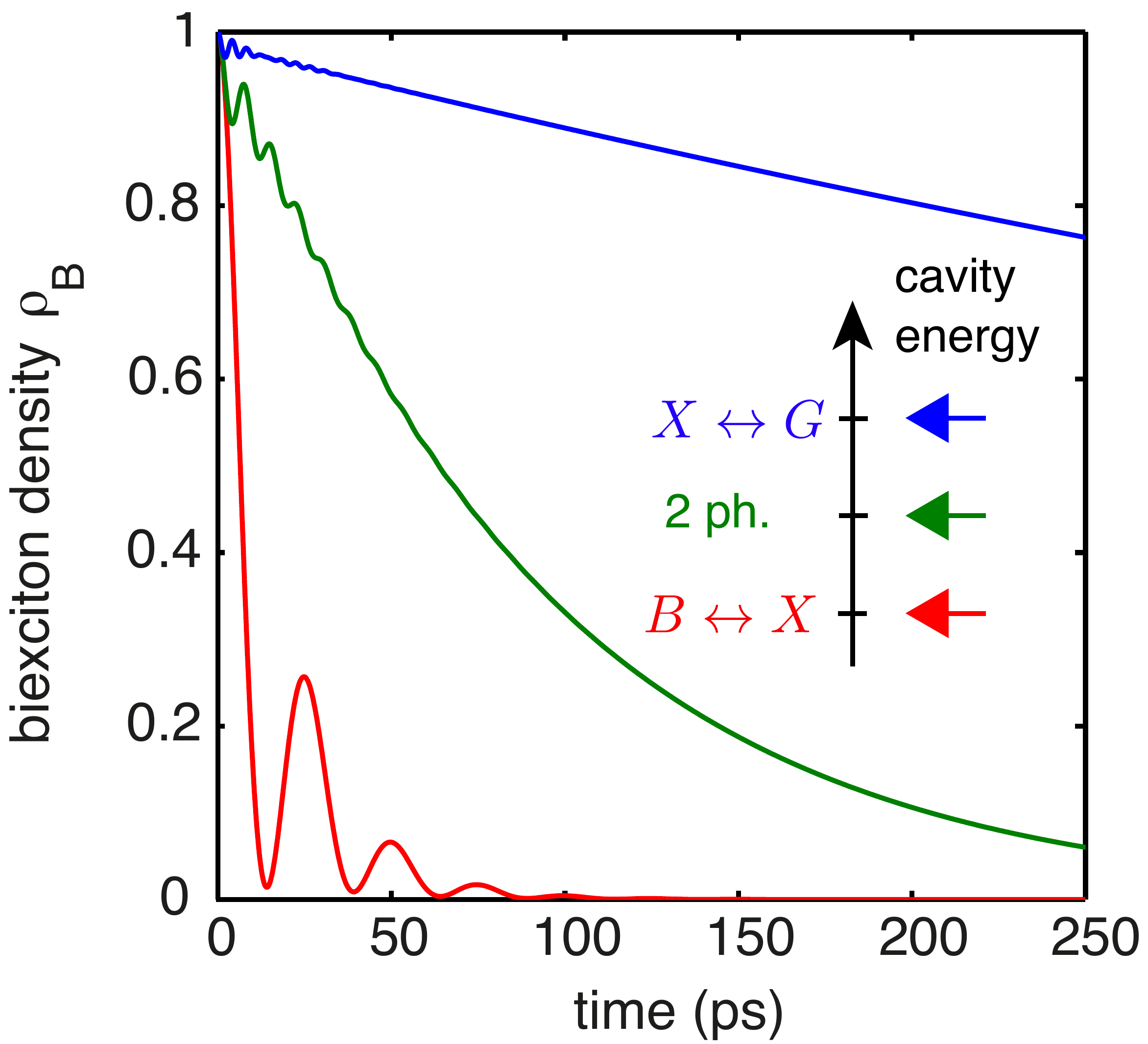}
\caption{Biexciton decay for different cavity frequencies. Results are shown for the cavity tuned to the biexciton to exciton transition (red line), to the exciton to ground state transition (blue line), and to the degenerate biexciton to ground state two-photon transition (green line), respectively. \textcolor{black}{Choosing to enhance one transition suppresses the other, because the transitions are seperated by the biexciton binding energy. The direct two-photon transition only occurs if the cavity mode is tuned accordingly, i.e. to half the biexciton energy.}}
\label{Einzeitig_N_B}
\end{figure}

To analyze the decay from the biexciton configuration including coupling to the system environment we use a density matrix theory. In this approach the system dynamic is given by the following Master equation:
\begin{equation}
\frac{\partial}{\partial t} \rho_s = - \frac{i}{\hbar} [H,\rho_s]+  \mathcal{L} _{\mathrm{cavity}} (\rho_s)+  \mathcal{L}_{\mathrm{pure}}  (\rho_s ) 
+ \mathcal{L}_{\mathrm{phonon}}(\rho_s )\,.
\label{Master-Equation}
\end{equation}
The coupling of the system to its environment and corresponding dissipative losses are included via the Lindblad terms $\mathcal{L}_{i}(\rho_s)$.\cite{Lindblad1976} These include emission of photons from the cavity on a time scale $1/\kappa$,
\begin{equation}
\mathcal{L}_{\mathrm{cavity}}(\rho_s) = \frac{\kappa}{2} \sum_{i=H,V}(2 b_i\rho_s b_i^{\dagger} -  b_i^{\dagger} b_i \rho_s - \rho_s  b_i^{\dagger} b_i )\,.
\end{equation}

Through interaction with the environment, electronic coherences experience a loss of phase information which is called pure dephasing \cite{Schumacher2012,PhysRevB.74.235310}
\begin{equation}
\mathcal{L}_{\mathrm{pure}}(\rho_s ) = -\frac{1}{2} \sum_{\chi , \chi' ; \chi \neq \chi'} \gamma_{\mathrm{pure}}^{\chi, \chi'} \vert \chi \rangle \langle \chi \vert \rho_s \vert \chi' \rangle \langle \chi' \vert\,.
\end{equation}
At low temperatures, pure dephasing can be as low as a few $\mu eV / K$ \cite{PhysRevLett.103.087405}. Here we assume $\gamma_{\mathrm{pure}}^{\chi, \chi'} = 1 \mu eV / K \ T$ for all electronic coherences. In high-quality cavities emission into off-resonant photon modes is strongly reduced but can not be prevented completely.  \textcolor{black}{These radiative losses can be included} through a term giving rise to changes in electronic populations but not generating photons in the system cavity modes:
\begin{equation}
\mathcal{L}_{\mathrm{rad}}(\rho_s ) = -\frac{\gamma_{\mathrm{rad}} \langle  \textcolor{black}{ \mathcal{B} } \rangle^2}{2} \sum_{i=X_H, X_V}  \left( \mathcal{L}_{\vert G \rangle \langle i \vert } +\mathcal{L}_{\vert i \rangle \langle B \vert }  \right) (\rho_s),
\label{Leaky-Modes}
\end{equation}
with $\mathcal{L}_{\sigma}(\rho_s) = \left(2 \sigma\rho_s \sigma^{\dagger} -  \sigma^{\dagger} \sigma \rho_s - \rho_s  \sigma^{\dagger} \sigma \right)$.
The renormalization factor $ \langle  \textcolor{black}{ \mathcal{B} } \rangle^2$ accounts for the reduced radiative emission due to phonon interaction.\cite{PhysRevX.1.021009} The decay constant $\gamma_{\mathrm{rad}}$ can be on the order of a few $\mu eV$ \cite{PhysRevB.81.035302,PhysRevLett.103.087405}.   
 \textcolor{black}{For the results shown below we  have checked that radiative losses only slightly affects the quantum efficiency of the photon emission into the cavity.\cite{1367-2630-13-11-113014}}

In the numerical evaluation, we solve the master equation, Eq.~(\ref{Master-Equation}),  \textcolor{black}{(for the phonon-assisted part see Eq.~(\ref{Lindblad-Phonon}) below).}  \textcolor{black}{The initial condition is} that the system is in the biexciton state and the cavity is empty. This state can be experimentally prepared \textcolor{black}{ by non-degenerate two-color two-photon Rabi flopping\cite{Stufler2006,Boyle20102485} and it has recently been demonstrated that the influence of phonons can be beneficial to deterministically prepare the biexciton state.\cite{PhysRevB.91.161302}}  The Fock space $\vert \chi; n_V, n_H \rangle$ is spanned by the joint spaces of the electronic states $\chi \in \{ B, X_H, X_V, G \}$ and the space of the photons $\vert n_V, n_H \rangle$. As expected from energy conservation, in the numerical evaluation full convergence is obtained by inclusion of photon states with up to $n = 2$ photons per cavity mode. The expectation value of any operator is calculated by taking the trace with the density operator; for example the biexciton population is calculated as $\rho_B=  \langle \vert B \rangle \langle B \vert  \rangle = \mathrm{tr}(\rho_s \vert B \rangle \langle B \vert)$.\cite{1367-2630-15-3-035019} \textcolor{black}{If not otherwise noted, below we use the following system parameters:}  \textcolor{black}{$g=\hbar/10$ $ps$, $\kappa = g$, $E_{XX}^B =1$ $meV$, $T = 4$ $K$ and $\delta_{\text{cavity}} = 0$ $meV$.}  \textcolor{black}{Here, we define the cavity detuning $\delta_{\text{cavity}} = \hbar \omega -E_{\text{2ph}}$ with $E_{\text{2ph}}=E_{\text{B}}/2$ being the degenerate two-photon resonance.}  In Fig.~\ref{Einzeitig_N_B} results are shown for different frequencies of the cavity modes. Depending on the cavity frequency different decay channels from the biexciton are enhanced or suppressed, respectively. When the cavity is tuned to the biexciton-to-exciton transitions, pronounced Rabi oscillations are visible as the population relaxes to the exciton configurations. Further decay to the ground state is suppressed however (not shown in Fig.~2); with the finite biexciton binding energy of $1\,\mathrm{meV}$ the exciton-to-ground state transitions are off-resonant to the cavity modes. When the cavity is resonant with the exciton-to-ground state transitions, no resonant emission from the biexciton can occur such that the decay of biexciton population is very slow. Tuning the cavity to half the biexciton energy, resonant with the higher-order degenerate two-photon biexciton-to-ground state transition, efficient relaxation to the ground state is observed while emitting two photons at once (and without creating exciton population). Below we investigate this latter process in more detail and analyze the properties of the emitted photon pair.

\subsection[Phonon-mediated transitions]{Phonon-Mediated Transitions}
Phonon effects are suspected to have an important influence on the photons generated in a quantum-dot cavity system with sufficiently strong coupling. Here, we investigate near resonant excitation only such that in the self-assembled InAs/GaAs quantum dots the main contribution is expected to stem from interactions with LA phonons (LO phonons are sufficiently well separated in energy).\cite{PhysRevX.1.021009} In the following we also include this coupling to take into account phonon-assisted cavity feeding.\cite{PhysRevB.81.155303} To include these interactions  we follow the analysis by Roy and Hughes for a two-level system coupled to a quantized optical mode\cite{PhysRevB.85.115309,PhysRevX.1.021009}  and extend it to a four-level system\cite{PhysRevB.93.115308,PhysRevB.81.155303} with two cavity modes as described in Sec.~\ref{System} above. 
 
\begin{figure}[t]
\centering
\includegraphics[scale=0.27]{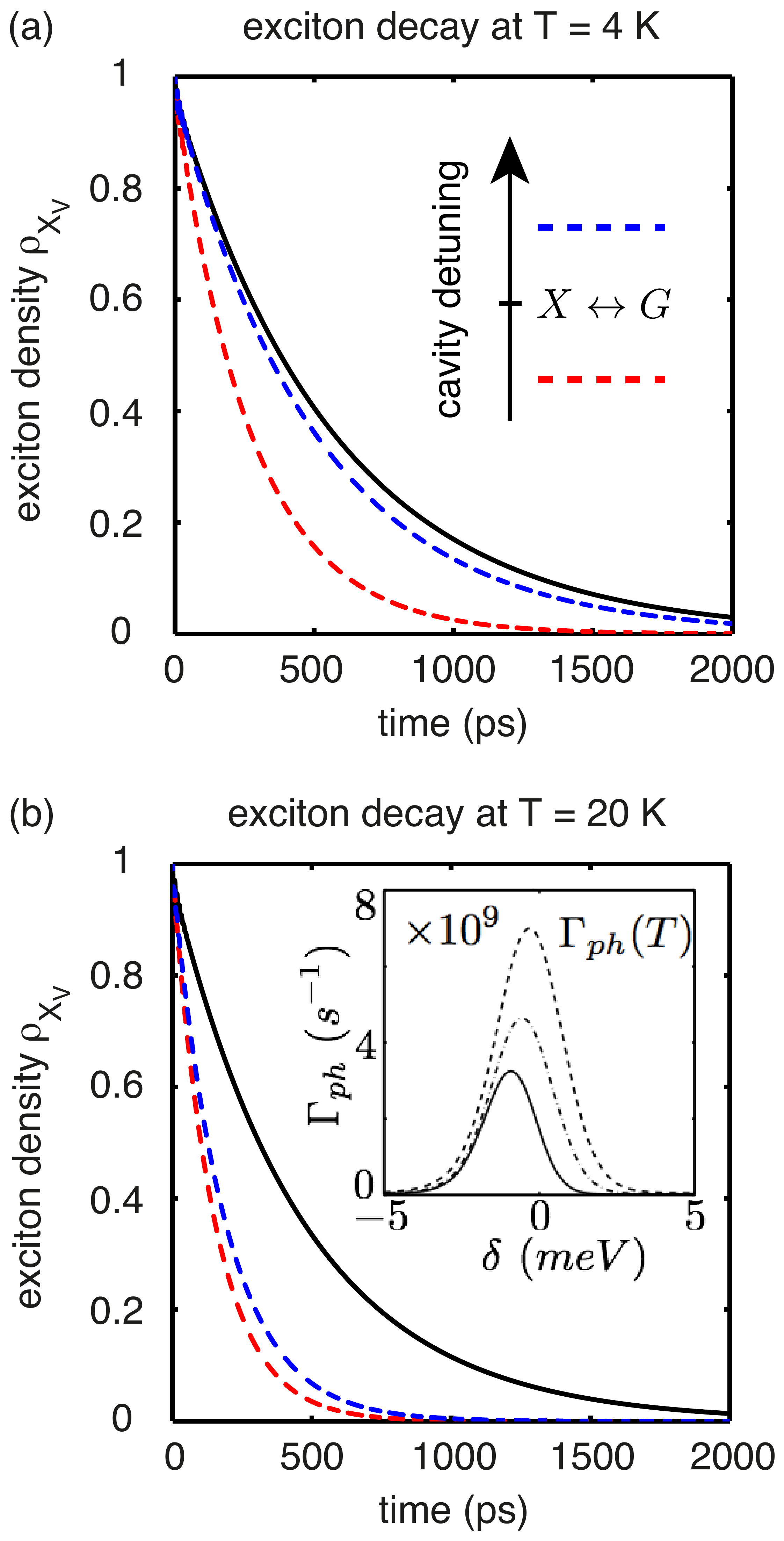}
\caption{\textcolor{black}{Phonon effect on exciton decay: a) Exciton decay at $4\,\mathrm{K}$ for cavity detunings of $\pm 500\,\mathrm{\mu eV}$ from the exciton to ground state transition. Without phonon-assisted transitions both detunings result in the same temporal decay (solid black line). If, however, phonon-assisted processes are included (dashed lines) even with a finite cavity detuning a quicker decay of the exciton is observed, \textcolor{black}{in agreement with previous measurements of cavity polariton life times.\cite{PhysRevX.5.031006}} Tuning the cavity to the red (red dashed line) results in a faster decay than in the case of the blue-detuned cavity (blue dashed line). b) Exciton decay as in b) but at $20\,\mathrm{K}$. Other parameters and line colors are the same as in a). The inset shows the phonon mediated transition rates $\Gamma(\delta)$ for  $\delta=\hbar\omega_i-E_i$ with $i=V,H$ and temperatures $T = 4\,\mathrm{K}$ (solid), $10\,\mathrm{K}$ (dot-dashed) and $20\,\mathrm{K}$(dashed). At low temperatures the phonon assisted cavity feeding process prefers a red detuned cavity, whereas at higher temperatures \textcolor{black}{it becomes} more symmetric and increases in strength.}}
\label{Gamma-ph_Ex-decay}
\end{figure}

The Hamiltonian including the LA phonons and electron-phonon interaction is then given by
\begin{eqnarray}
H &=& H_S+H_{\mathrm{phonon}}+H_{\mathrm{QD-phonon}} \nonumber  \\
  &=& H_S+\sum_{q} \hbar \omega_q a_q^{\dagger} a_q + \sum_{i,q} \vert \chi_i\rangle \langle \chi_i \vert\lambda_q^i ( a_q^{\dagger} + a_q)\,,
\end{eqnarray}
with $H_S$ from Eq.~(\ref{Hamiltonian}), and electronic configurations $\chi_i = \{ X_V, X_H, B \}$ and phonon creation (annihilation) operators $a_q^{\dagger} $ ($a_q$).  \textcolor{black}{The energy of the phonons in mode $q$ is $ \hbar \omega_q$ and the coupling to the quantum dot state $i$ is given by $\lambda_q^i$.}
By transformation into the polaron frame the explicit appearance of the phonons can be removed from the Hamiltonian.\cite{Florian2013,PhysRevX.1.021009,PhysRevB.81.155303}  \textcolor{black}{The transformation into the polaron frame is done following Ref.~\onlinecite{PhysRevB.93.115308}} with $H' = e^A \ H \ e^{-A}$ and
\begin{equation}
 A = \sum_{i,q} \frac{\lambda_q^i }{\omega_q} \vert \chi_i\rangle \langle \chi_i \vert ( a_q^{\dagger} - a_q).
 \end{equation}
 \textcolor{black}{The transformed quantum-dot cavity Hamiltonian then reads } 
 \begin{eqnarray}
 H'  &=&  \sum_{i} \tilde{E_i}\vert \chi_i\rangle \langle \chi_i \vert + \sum_{j}  \hbar\omega_j b_j^{\dagger}b_j + \langle \mathcal{B} \rangle X_g  \nonumber  \\
&+&  \sum_{q} \hbar\omega_q a_q^{\dagger} a_q + \zeta_g X_g -\zeta_u X_u.
\end{eqnarray}
\textcolor{black}{ The cavity  and quantum - dot part of} this Hamiltonian has the same structure as Eq.~(\ref{Hamiltonian}), but with a renormalized electron-photon coupling constant $g \rightarrow g \langle \textcolor{black}{ \mathcal{B} }\rangle$.\cite{PhysRevB.65.235311}   \textcolor{black}{The polaron shift is assumed to be included in the electronic energies given in the Hamiltonian with $\tilde{E_i} = E_i -\frac{{\lambda_q^i}^2 }{\omega_q}$,} and $\langle \textcolor{black}{ \mathcal{B} } \rangle$ is the thermal average of the phonon-bath displacement \cite{PhysRevX.1.021009}
\begin{equation}
\langle \textcolor{black}{  \mathcal{B} } \rangle = \exp \left( -\frac{1}{2} \int_0^{\infty} d \omega \frac{J(\omega)}{\omega^2} \coth(\frac{\beta \hbar \omega}{2})  \right) .
\label{Phonon-Bath-Displacement}
\end{equation}
The temperature dependence is included by $\beta=\frac{1}{k_b T}$. 
The spectral function $J(\omega)$ that describes the interaction between the electrons in the quantum-dot and the acoustic phonons coupling via a deformation potential is given by
\begin{equation}
J(\omega)=\sum_q \lambda^2_q \delta(\omega-\omega_q) =\alpha_p \omega^3 e^{-\frac{\omega^2}{2\omega_b^2}}
\end{equation}
with  $\alpha_p=0.06\,\mathrm{ps^2}$ and $\omega_b=1\,\mathrm{meV}$ for InAs/GaAs quantum-dots.\cite{PhysRevX.1.021009,PhysRevB.85.115309}  \textcolor{black}{The biexciton is assumed to couple twice with the phonon bath compared to the excitons with $\lambda^B_q = 2 \lambda^{X_V}_q =2 \lambda^{X_H}_q =2 \lambda_q$.\cite{PhysRevB.93.115308}} \textcolor{black}{Additionally, $H_{\text{qd - ph}}' =  \zeta_g X_g -\zeta_u X_u$ is the transformed quantum dot - phonon bath Hamiltonian. Here  $\zeta_g = \frac{1}{2}(\mathcal{B}_{+} + \mathcal{B}_{-} - 2\langle \mathcal{B} \rangle)$,  $\zeta_u = \frac{1}{2i}(\mathcal{B}_{+} + \mathcal{B}_{-})$ with $\mathcal{B}_{\pm} = \exp(\pm \sum_q \frac{\lambda_q}{\omega_q}(a_q - a^{\dagger}_q))$ and $\langle \mathcal{B}_{\pm} \rangle = \langle \mathcal{B} \rangle$, as well as  $X_g = \mathcal{X}+h.c.$, $X_u = i(\mathcal{X}-h.c.)$.\cite{PhysRevX.1.021009}}

\begin{figure}[t]
\centering
\includegraphics[scale=0.19]{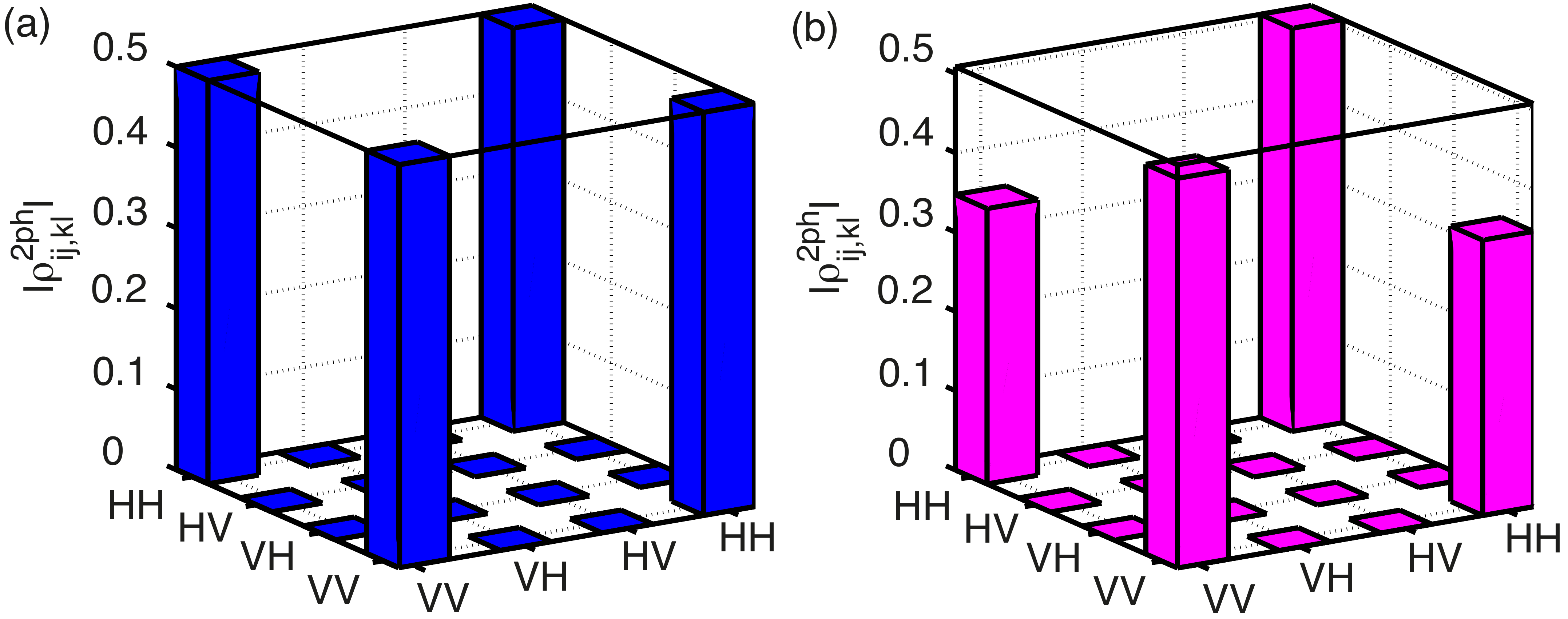}
\caption{ \textcolor{black}{The two-photon density matrix is shown for a) the maximally entangled state and b) for an arbitrarily entangled state. A finite fine structure splitting leads to destructive interference of signals in the two different decay paths, decreasing the non-zero off-diagonal elements. This decrease is directly related to a reduced degree of polarization entanglement of the emitted photons.}}
\label{2ph-densitymatrix-comp}
\end{figure}

By treating the phonons in a second order Born-Markov approximation in the master equation \textcolor{black}{and tracing out the phonon bath degrees of freedom}, the LA-phonon interaction is now included via the extra Lindblad term in Eq.~(\ref{Master-Equation}) with\cite{Ulhaq:13,PhysRevB.85.115309,PhysRevX.1.021009,PhysRevLett.106.247403}
\begin{multline}
 \mathcal{L}_{\mathrm{phonon}}(\rho_s ) =   \\ - \frac{1}{\hbar^2} \sum_{i = g,u}\ \int_0^t d\tau \ ( X_i(t)X_i(\tau)\rho_s(t) G_m(t-\tau)    \\ 
- X_i(\tau)\rho_s(t)X_i(t)G_m(t-\tau) + h.c.) \,,
\label{Lindblad-Phonon} 
\end{multline}
 \textcolor{black}{with $G_g(t) = \langle \mathcal{B}\rangle^2(\cosh(\phi(t)) -1)$ and $G_u(t) = \langle \mathcal{B}\rangle^2 \sinh(\phi(t)) $.\cite{PhysRevB.85.115309} The evaluation of all dynamical quantities is done in the interaction picture following Ref.~\onlinecite{Breddermann2016}} 
 
\textcolor{black}{In the case of an optical transition from the exciton to its ground state the rates for phonon mediated transitions derived in Ref.~\onlinecite{PhysRevX.1.021009} are obtained as shown in the inset of Fig.~3(b):} 
\begin{equation}
\Gamma_{\mathrm{ph}} = 2 \langle  \textcolor{black}{ \mathcal{B} } \rangle {}^2 g^2 \Re \left[ \int_0^{\infty} d\tau e^{\pm i\Delta \tau} \left(  e^{\phi(\tau)} -1 \right)  \right] .
\label{Phonon-rates}
\end{equation}
\textcolor{black}{Here $\Delta=\hbar \omega_{i}-E_{\text{i}}$ is the energy difference of the corresponding cavity-mode $\hbar \omega_i$ and electronic transition energies $E_{\text{i}}$ between exciton and ground state.}  The phonon correlation functions $\phi(t)$ are given by\cite{PhysRevX.1.021009} 
\begin{equation}
\phi(t) = \int_0^{\infty} d \omega \frac{J(\omega)}{\omega^2}  \left( \coth(\frac{\beta \hbar \omega}{2}) \cos(\omega t) -i \sin(\omega t) \right).
\end{equation}

The effect of the phonon-assisted processes on the decay of an exciton is shown in Figs.~\ref{Gamma-ph_Ex-decay}(a) and (b) for temperatures of $ T = 4$ K and $T=20$ K. Figure~3 shows that even for an off-resonant cavity mode, faster decay of exciton populations is found with increasing temperature by the increased interaction with the phonon bath. Also, at relatively low temperatures, a cavity detuned to the  \textcolor{black}{red} leads to a faster decay of the exciton densities compared to the  \textcolor{black}{blue}-detuned case. This is because the phonon absorption from the phonon bath (needed to assist the exciton decay for blue-detuned cavity) is strongly reduced at low temperatures. This asymmetry is also visible in the inset. The lower the temperature the more pronounced this asymmetry is. Already at $T=20 \ K$, phonon assisted processes involving phonon absorption and emission are almost balanced such that the decay of exciton population at $T=20 \ K$ in Fig.~\ref{Gamma-ph_Ex-decay} $b)$ is almost the same for postive and negative detuning. At $T=4 \ K$  in Fig.~\ref{Gamma-ph_Ex-decay} $a)$, the role of phonon-assisted processes is only weak when phonon absorption is required for emission into a cavity mode that is tuned to the blue.

\subsection{The Two-Photon Density Matrix}

To know the state of the emitted photons the decay path from the biexciton to the ground state must be known. In general, it is possible that the biexciton decays purely via a single cavity mode by emitting two photons of the same polarization such that the photons are in the state $\vert VV \rangle$ or $\vert HH \rangle$ or the biexciton can decay by emitting two photons of different polarization, resulting in a $\vert VH \rangle$ or $\vert HV \rangle$ state. The two-photon density matrix contains full information about the quantum state of the two emitted photons and is obtained experimentally by quantum-state tomography based on photon correlation measurements.\cite{JJAP.46.7175} The two-photon density matrix is  calculated as the double time integral
\begin{equation}
\rho_{ij,kl}^{2 \mathrm{ph}} = \iint G_{ij,kl}^{(2)}(t,\tau) \  dt d\tau\,,
\label{two-photon-density-matrix}
\end{equation} 
of the second order photon autocorrelation function
\begin{eqnarray}
G_{ij,kl}^{(2)}(t,\tau) & = & \langle  b_i^{\dagger}(t) b_j^{\dagger}(t+ \tau) b_k(t+\tau) b_l(t) \rangle \nonumber \\
& = & \mathrm{tr} ( \rho_s  b_i^{\dagger}(t) b_j^{\dagger}(t+ \tau) b_k(t+\tau) b_l(t) )\,.
\label{G2-t-tau}
\end{eqnarray}
We use the quantum regression theorem to calculate these two-time expectation values.\cite{Carmichaelbook} The diagonal elements of (\ref{G2-t-tau}) contain information about the photon statistics where the off-diagonal elements contain information about the polarization entanglement of the two photons. The two-photon density matrix fulfils $\mathrm{tr}(\rho^{2 \mathrm{ph}}) =1$. In the system studied here, the two-photon density matrix only contains up to four non-zero matrix elements (cf. Fig.~4) and can be simplified accordingly to $\rho_{i,j}=\rho_{ii,jj}^{2 \mathrm{ph}}$, with $i,j\in\{H,V\}$. In this case the degree of polarization entanglement can be measured by the concurrence\cite{RevModPhys.81.865}
\begin{equation}
C = 2 \vert \rho_{H,V} \vert \,.
\label{Concurrence}
\end{equation}
Figure \ref{2ph-densitymatrix-comp} shows the two-photon density matrix for a maximally entangled state and a state resulting from the emission of a quantum-dot cavity system with finite fine structure splitting of the two exciton states. In the latter case the decay is favored through the \textcolor{black}{$X_H$} exciton over the $X_V$ exciton, which results in a slight increase of the \textcolor{black}{$\rho_{H,H}$} contribution.\ Also as a result of the exciton splitting, the "which-path" information for the biexciton decay is revealed such that the off-diagonal elements and with them the concurrence (as a measure of polarization entanglement) is reduced.

To gain insight into the statistics of emitted photons, one has to consider the time integrated $g^2$ function
\begin{equation}
g_{i,i}^2(\tau) := \int G_{ii,ii}^{(2)}(t,\tau) \  dt\,.
\label{g2-tau}
\end{equation}
This photon correlation function gives the probability to detect another photon with delay $\tau$ after the first photon was detected and is a useful measure to evaluate the statistical properties of the emitted photons.

\begin{figure} [t]
\centering
\includegraphics[scale=0.35]{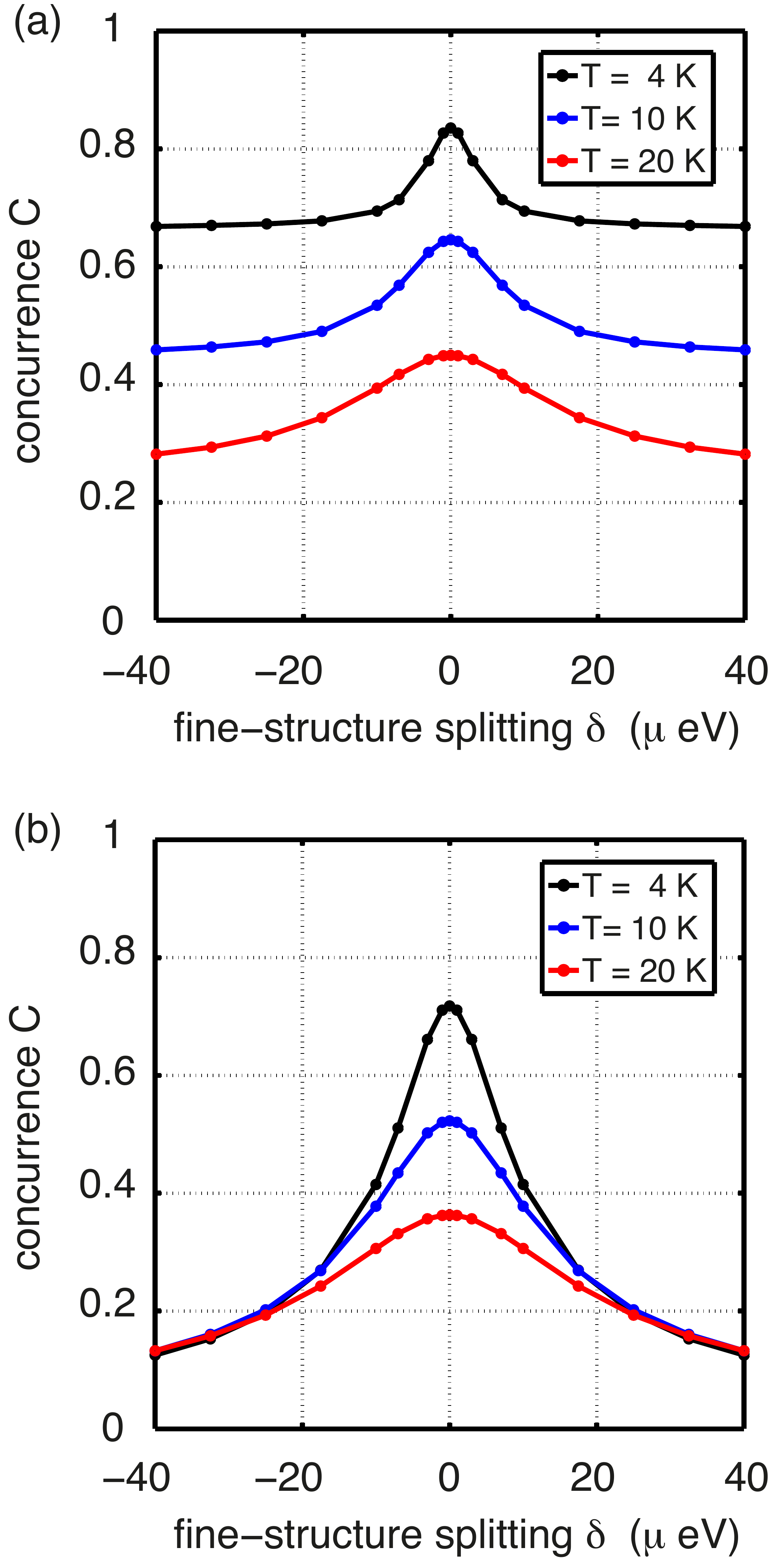}
\caption{\textcolor{black}{Degree of polarization entanglement for various  values of excitonic fine-structure splitting and different temperatures. Panel a) depicts the case of a high quality cavity with $g / \kappa = 1$ in which two photons are emitted at the same time. This results in a high degree of entanglement that is insensitive to the fine structure splitting  \textcolor{black}{at low temperatures}.
In b) the cavity parameter is chosen to be $g / \kappa = 0.02$ such that the usual properties reported for cascaded emission are observed. }}
\label{Conc_T_kappa}
\end{figure}

\section{Results \& Discussion}

In this section we discuss the main results obtained for the photons emitted from a quantum dot biexciton when embedded inside a cavity with its resonance tuned close to half the biexciton energy. In Sec.~\ref{Section-Concurrence} polarization entanglement is discussed \textcolor{black}{for the ideal and several non-ideal cases,} and in Secs.~\ref{Section-Photon-Statistics} and \ref{Section-Photon-Spectra} photon statistics and spectral properties, respectively.

\subsection{Polarization Entanglement}
\label{Section-Concurrence}

\begin{figure}[t]
\centering
\includegraphics[scale=0.35]{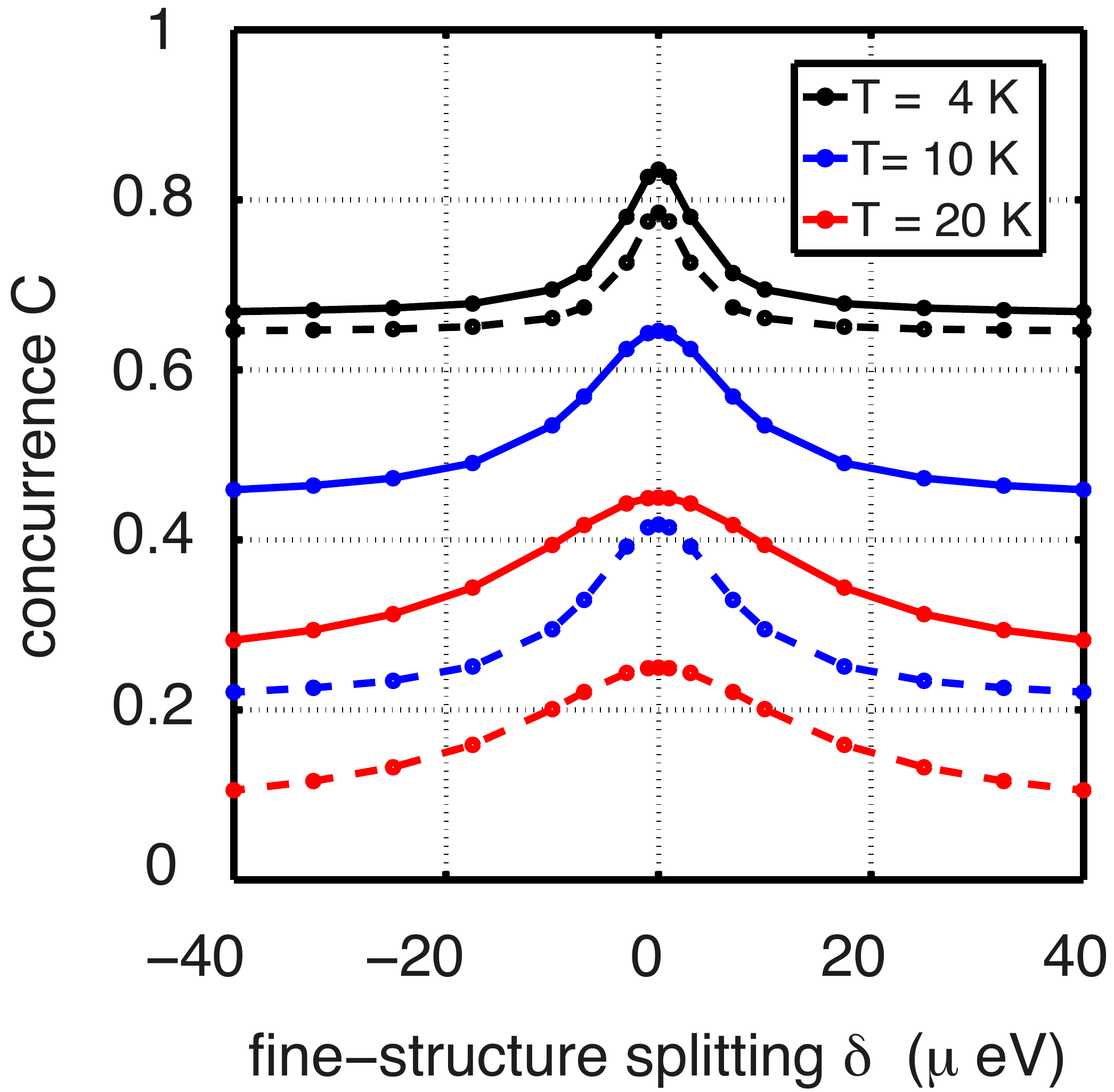}
\caption{Concurrence for different temperatures dependent on the electronic fine-structure splitting. The solid line shows the result for a biexciton binding-energy of $E_{XX}^B = 1\,\mathrm{meV}$ and the dashed line is for $E_{XX}^B = 3\,\mathrm{meV}$.}
\label{Conc-EXXB-VGL}
\end{figure}

The polarization-entanglement of the emitted photons depends on various system parameters. In order to give a detailed picture, the dependence on fine-structure splitting, temperature, biexciton binding energy, different loss mechanisms, cavity quality, and cavity detuning are discussed in the following. The concurrence, Eq.~(\ref{Concurrence}), is used to quantify the degree of entanglement. The calculated concurrence as a function of the excitonic fine-structure splitting is shown in Fig.~\ref{Conc_T_kappa} for different values of  cavity-quality and temperature. The cavity quality determines the cavity enhancement of the direct two-photon emission process over the cascaded decay. For a sufficiently high cavity quality, the two-photon process dominates the emission from the biexciton such that no "which-path" information is revealed and a consequently a high degree of entanglement is achieved that is insensitive to exciton fine structure splitting. For a lower quality cavity the biexciton mostly decays through the biexciton-exciton cascade such that the usual sensitivity of the polarization entanglement on the fine structure splitting is recovered.\cite{1367-2630-9-9-315} Importantly, at low temperatures and high quality cavity, we obtain high degrees of polarization entanglement very similar to the case where phonon-assisted cavity feeding is neglected.\cite{Schumacher2012} With the cavity tuned near half the biexciton energy the biexciton to exciton transition is red-shifted from the cavity mode. Thus at low temperatures phonon-assisted cavity feeding leading to a decay through the cascade is  suppressed as the probability for phonon absorption from the phonon bath is very low. With increasing temperature the phonon bath population increases and polarization entanglement is reduced for both high and low quality cavity. 

The biexciton binding energy is one of the quantum dot's intrinsic properties, which can be modified by, e.g., material composition, growth conditions, or post growth by applying strain  \textcolor{black}{and electrical fields}\cite{PhysRevLett.109.147401,Juska2013}. The dependence of the polarization entanglement on the biexciton binding energy is shown in Fig.~\ref{Conc-EXXB-VGL}. \textcolor{black}{At low temperature and for the parameters studied here we find that the concurrence slightly decreases with increasing biexciton binding energy. The increase in biexciton binding energy also leads to an increased detuning of the degenerate two-photon transition from the single-photon resonances, which comes at the loss of the resonance enhancement of the two-photon process from the near-by single-photon transitions. This leads to an overall slower decay of the biexciton. Consequently, with increasing temperature when phonon-assisted transitions more efficiently feed the cascaded decay, an increased biexciton binding energy is actually found to reduce the polarization entanglement of the emitted photons even more than at low temperatures.}  

\begin{figure}[t]
\centering
\includegraphics[scale=0.3]{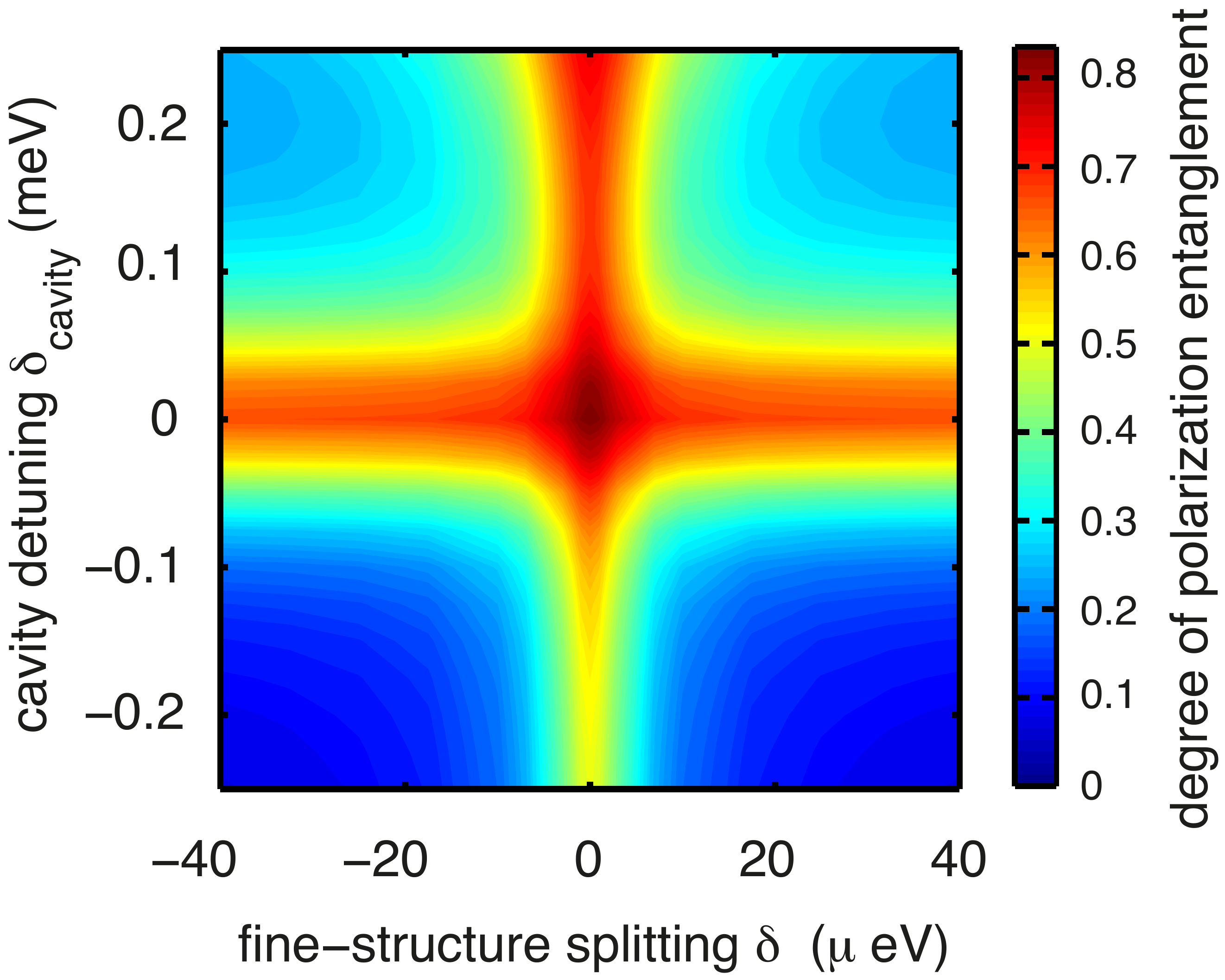}
\caption{The figure shows the dependence of the concurrence on the cavity detuning at a fixed temperature $T=4\,\mathrm{K}$ \textcolor{black}{ and with a high-quality cavity. At zero cavity detuning two degenerate photons are emitted simultaneously yielding a high degree of polarization entanglement. In the case of a positive cavity detuning the cavity is shifted towards the exciton to ground state transition and the concurrence resembles the usual cascaded emission feature. A negative cavity detuning favors the biexciton to exciton decay, also hindering the two photon emission.}}
\label{Conc_Scan_dcav}
\end{figure}

In addition to the electronic properties of the quantum-dot discussed above, for the emission scheme discussed here, tuning the optical cavity near the two-photon resonance is very important for generating highly entangled photon pairs. Figure~\ref{Conc_Scan_dcav} shows the concurrence for different detunings $\delta_{\mathrm{cavity}}$ of the cavity mode from the two-photon resonance condition. The highest concurrence is obtained for $\delta_{\mathrm{cavity}}\approx0$, on resonance with the two-photon emission process. If the cavity energy is detuned from this ideal condition, the cascaded decay takes over the emission dynamics and the concurrence is reduced accordingly. We note that the asymmetry in Fig.~\ref{Conc_Scan_dcav} for positive and negative detuning is caused by the near-by single-photon resonances of the cascade.

\begin{figure}[t]
\centering 
\includegraphics[scale=0.35]{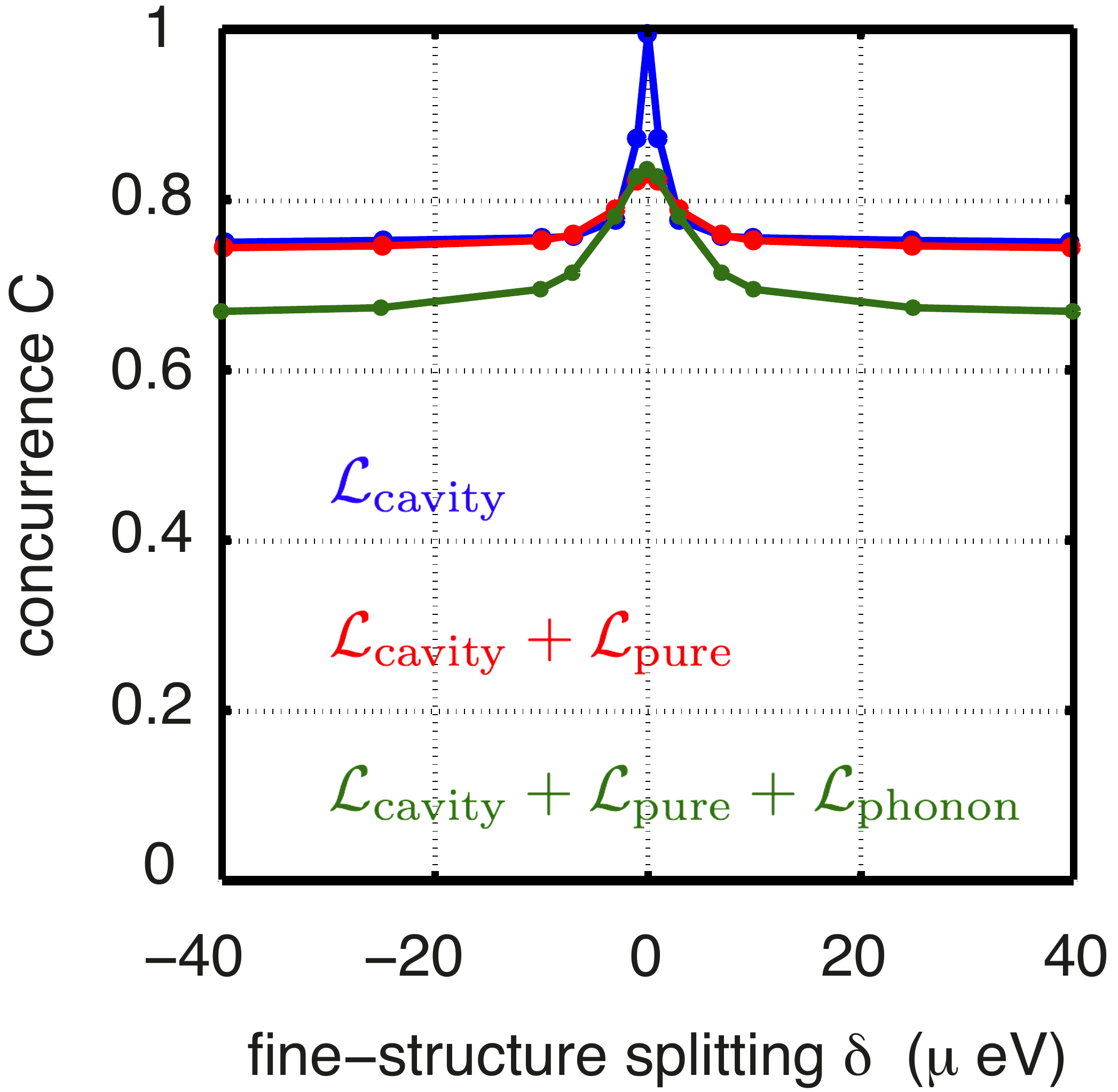}
\caption{Influence of decay mechanisms on the concurrence at T $= 4\,\mathrm{K}$. \textcolor{black}{ If only the loss of photons from the cavity is active (blue line) the concurrence reaches its theoretical maximum of unity for zero fine-structure splitting. An additional dephasing mechanism, such as pure dephasing (red line), reduces the maximal possible degree of entanglement. In the case of phonon-assisted cavity feeding (green line) this effect is more pronounced.}}
\label{Conc-Decay-mechanism}
\end{figure}

In the remainder of this section we would like to further discuss the role of the different loss mechanisms included in the calculations discussed above. Figure~\ref{Conc-Decay-mechanism} shows the concurrence when the different loss mechanisms are selectively switched off. The most fundamental loss is caused by the loss of photons from the cavity. Only including this mechanism, a maximally entangled state is obtained at zero fine-structure splitting. Loss of electronic coherence (pure dephasing) limits the maximally achievable entanglement also at zero fine-structure splitting. A further overall reduction of the concurrence is caused by the coupling to the phonon bath. \textcolor{black}{At low temperatures,} this effect causes a nearly constant offset on the energy range of $ \pm 40 \,\mathrm{\mu eV}$ considered here. \textcolor{black}{The peak in the concurrence at zero fine-structure splitting is caused by the overlap of the excitons with a line with given by the pure dephasing. With increasing temperature this peak is broadened by the phonon contributions.} A further loss mechanism not included in the results discussed above is caused by the emission of photons into optical modes other than those cavity modes explicitly considered part of the system. Emission into these leaky modes mostly reduces the total brightness of the quantum-dot cavity system as a source of entangled photon pairs. If loss into leaky modes through Eq.~(5) is explicitly included in our calculations it generally slightly increases the concurrence of those photons emitted from the system cavity modes. Other than that the concurrence shows the same dependence on system parameters as in the scenario with no radiative decay present.  Finally, we would like to note, that in all scenarios studied, polarization entanglement can be further increased by spectral filtering such that only those photons from the direct two-photon transition are detected.

\begin{figure}[t]
\centering
\includegraphics[scale=0.32]{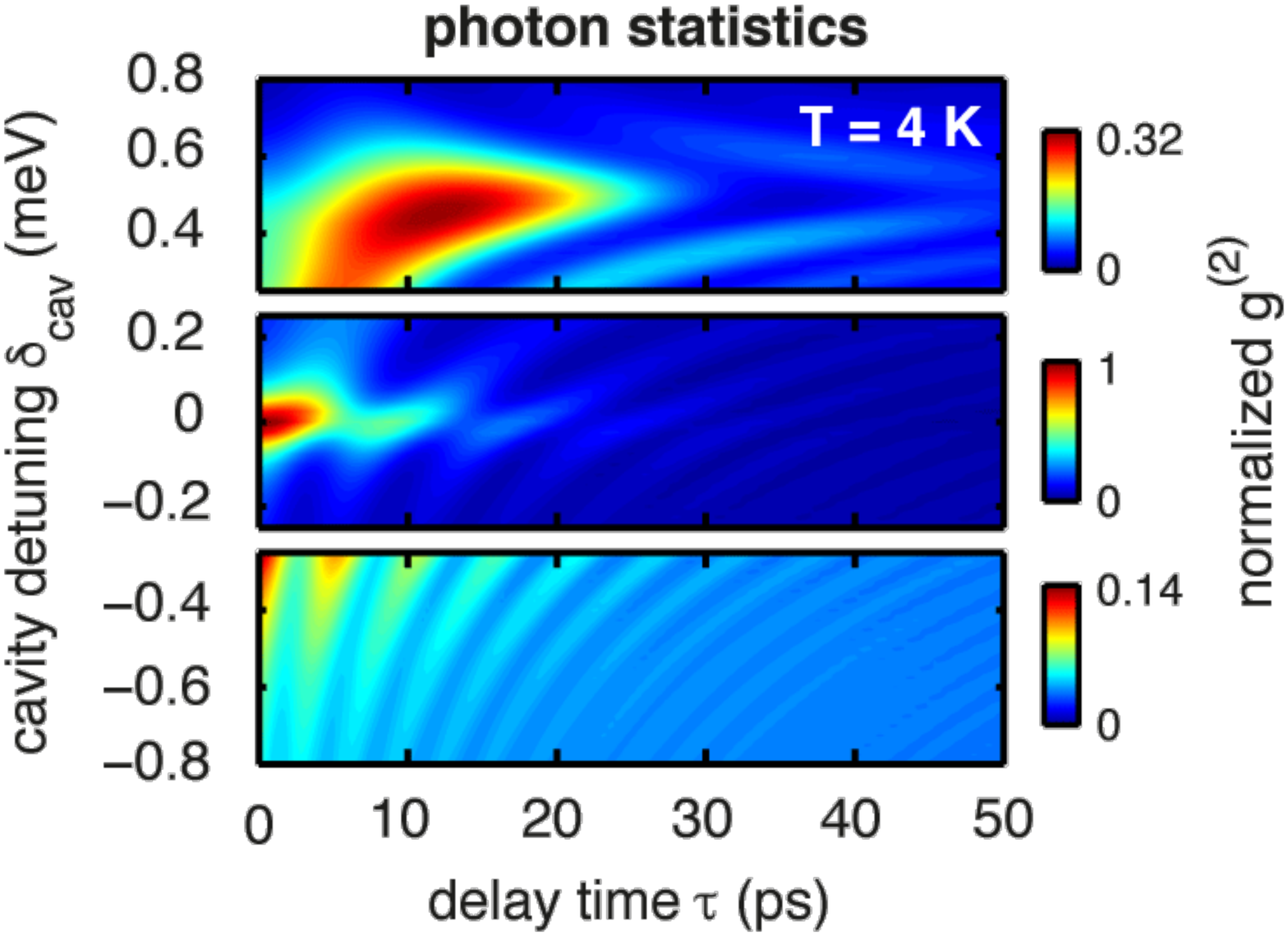}
\caption{ \textcolor{black}{Photon statistics at $T= 4\,\mathrm{K}$. The influence of the LA-phonons on the photon statistics given by the $g^2$ -  function is shown. Note that each panel has an individual color bar. In the top panel the anti-bunching of photons is visible if the cavity is tuned to the exciton at $\delta_{\text{cavity}} = 0.5\,\mathrm{meV}$. Only after the first photon was emitted from the biexciton, the second photon is emitted resonantly into the cavity with the characteristic delay of the cascaded emission. If, as depicted in the middle panel, the cavity is at resonance with the degenerate two-photon emission process photon bunching can be observed.  In the bottom panel the cavity is tuned close to the biexciton to exciton transition at  $\delta_{cavity} = -0.5\,\mathrm{meV}$. Here, the second photon from the exciton to the ground state transition is emitted into a red tuned cavity which benefits from phonon interaction such that a prolonged two photon correlation is observed.}}
\label{g2_scan_dcav_kappa=1_T=4K_2}
\end{figure}

\subsection{Photon Statistics}
\label{Section-Photon-Statistics}

\begin{figure}[t]
\centering
\includegraphics[scale=0.28]{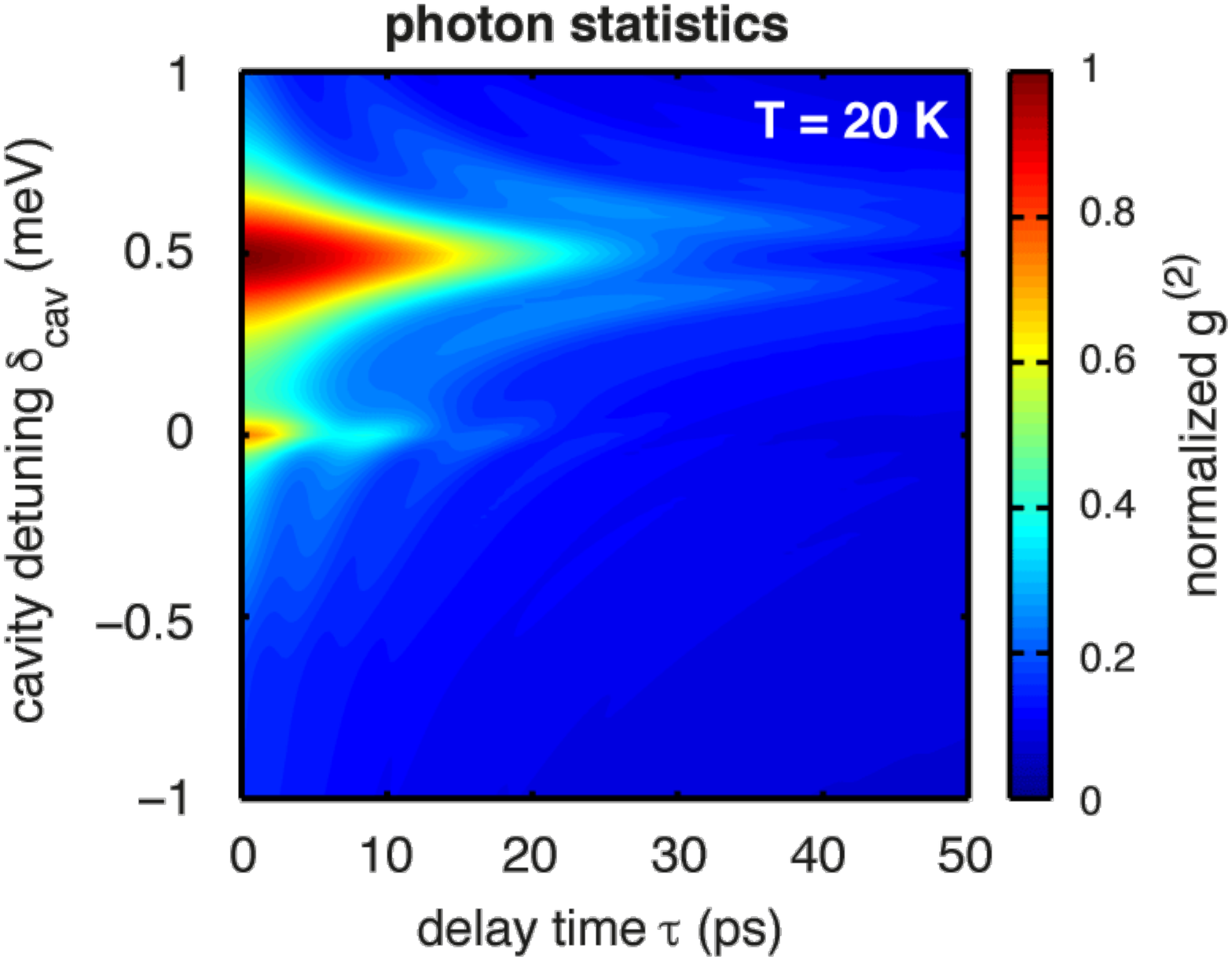}
\caption{  \textcolor{black}{Photon statistics at $T= 20\,\mathrm{K}$: At higher temperatures compared to the case shown in Fig.~\ref{g2_scan_dcav_kappa=1_T=4K_2} the photon anti-bunching feature is not present in the photons statistics if the cavity is tuned to the exciton to ground state transition $\delta_{\text{cavity}} = 0.5\,\mathrm{meV}$.}}
\label{g2_scan_dcav_kappa=1_T=20K_2}
\end{figure}

The photon statistics reveal information about the temporal emission properties in a photon mode, here the $H$ mode. The photon statistics are calculated as the time-integrated second order correlation function given in Eq.~(\ref{g2-tau}). The computed \textcolor{black}{ results of $g^2(\tau)$ are shown in Figs.~\ref{g2_scan_dcav_kappa=1_T=4K_2} and \ref{g2_scan_dcav_kappa=1_T=20K_2} for two different temperatures and for varying detunings of the cavity from the two-photon resonance}. The fine-structure splitting is set to $0\,\mathrm{meV}$. We find that for low temperature, \textcolor{black}{ Fig.~\ref{g2_scan_dcav_kappa=1_T=4K_2}}, a clear anti-bunching of the photons is observed \textcolor{black}{ if the cavity is resonant with the exciton to ground state transition $\delta_{\mathrm{cavity}}=+0.5 \,\mathrm{meV}$}. In this case the photon density inside the cavity is comparatively small at all times \textcolor{black}{ and the biexciton to exciton transition is far off-resonant and consequenctly inefficient  because the first photon is emitted into a blue-detuned cavity mode.} \textcolor{black}{Thus, these photon pairs }show the characteristic anti-bunching of the cascaded decay \textcolor{black}{at low temperatures}. If the biexciton to exciton transition is resonant with the cavity mode at $\delta_{\mathrm{cavity}}=-0.5\,\mathrm{meV}$, the transition to the ground state is   \textcolor{black}{suppressed } by the exciton binding energy of $ 1\,\mathrm{meV}$  \textcolor{black}{ and consequently} the second photon is emitted into a red-detuned cavity mode, which benefits from phonon assisted cavity feeding.   \textcolor{black}{Here,} a slowly decaying $g^2(\tau)$ is observed for long delays $\tau$, \textcolor{black}{with the maximum at $\tau = 0\,\mathrm{ps}$}. If the cavity is resonant with the two-photon transition at $\delta_{\mathrm{cavity}}= 0\,\mathrm{meV}$, the photon emission is bunched \textcolor{black}{since} the two photons are generated at the same time. At higher temperature, $20\,\mathrm{K}$ \textcolor{black}{ in Fig.~\ref{g2_scan_dcav_kappa=1_T=20K_2}}, features are smeared out as for all detunings phonon-assisted transitions are enabled, leading to more complex decay dynamics. The asymmetry caused by different phonon absorption and emission rates is clearly visible.  \textcolor{black}{Especially, in the case of $\delta_{\mathrm{cavity}}=+0.5\,\mathrm{meV}$, the first photon from the biexciton to the exciton is emitted faster into the cavity mode than at low temperatures. As a result a more bunching-like photon statistics can be observed. However, direct two photon emission is possible only if the cavity is tuned to the two photon resonance.}

\subsection{Emission Spectra}
\label{Section-Photon-Spectra}
\begin{figure}[t]
\centering
\includegraphics[scale=0.18]{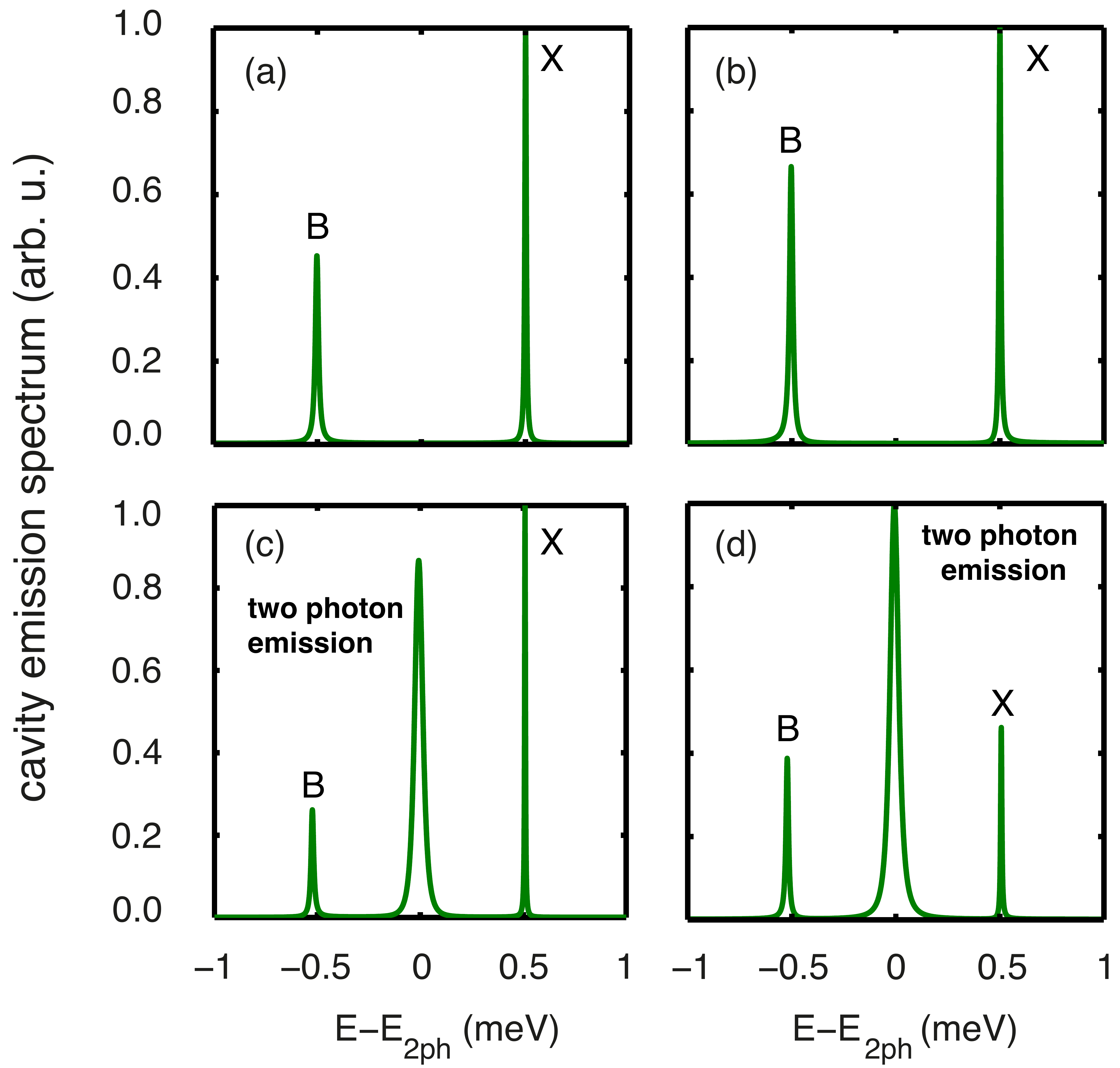}
\caption{\textcolor{black}{Emission spectra. The cavity modes are tuned to the two-photon resonance, $\delta_{\text{cavity}}=0$. The top row  shows the results for low quality cavity ($g/\kappa = 0.02$), without a) and with b) LA-phonon contributions. In the bottom row, the corresponding results for high-quality cavities ($g/\kappa = 1$) are shown in $c)$ and $d)$, respectively. LA-phonons feed the central emission line with photons originating from the exciton cascade and thus introduce a small proportion of photons that carry a "which-path" information at the direct two-photon emission energy.}}
\label{Emissions_Spectra_w_wo_G_Phonon}
\end{figure}

Below we discuss the spectral properties of the emitted photons. To analyze the spectral shape of the emitted photons the physical cavity emission spectrum\cite{Eberly:77,1367-2630-13-11-113014,Breddermann2016} $S_{C}(\omega)$ is calculated from the two-time photon correlation function as 
\begin{equation}
S_{C}(\omega) = \Re \int_0^{T} dt \, \int_0^{T-t}  d \tau \, \langle  b_i^{\dagger}(t) b_i(t+\tau) \rangle e^{i \omega \tau}\,.
\label{Cavity-spectrum}
\end{equation}
The photon correlation function is calculated using the quantum regression theorem,\cite{Carmichaelbook} with $T$ sufficiently large to obtain a time integrated spectrum after the system has fully relaxed to its ground state. The calculated spectra are shown in Fig.~\ref{Emissions_Spectra_w_wo_G_Phonon}. The fine-structure splitting is zero in these calculations, such that the emission in both cavity modes is identical. The cavity is tuned to the two-photon resonance \textcolor{black}{$\delta_{\text{cavity}}=0$} and the temperature is at $4\,\mathrm{K}$. For low cavity quality (upper row) the emission is only visible at the biexciton to exciton transistion \textcolor{black}{($E=E_{\text{2ph}}-0.5\,\mathrm{meV}$)} and at the ground state to exciton transition \textcolor{black}{($E=E_{\text{2ph}}+0.5\,\mathrm{meV}$)}. Including the coupling to the phonon bath, the emission lines are altered to a slightly broadened and slightly asymmetric shape. Without phonon coupling the areas under the two emission peaks from the cascade are identical. For a high-quality cavity, a third emission peak is visible at the cavity frequency. This peak mostly stems from the direct two-photon biexciton to ground state transition and gains additional smaller contributions from phonon-assisted cavity feeding when the coupling to the phonon bath is included. The two-photon emission line is slightly shifted from the ideal two-photon resonance condition because of the strong internal coupling between states.\cite{PhysRevB.81.035302} Even with finite fine structure splitting, the two photons emitted into the central peak are highly entangled in their polarization state. However, for finite fine-structure splitting, phonon-assisted emission slightly lowers this entanglement as it mixes in photons emitted from the cascade, e.g., through $b_{V}^{\dagger} |X_{V}\rangle\langle B|$ terms in the corresponding Lindblad term, Eq.~(\ref{Lindblad-Phonon}), which carries the path information of the decay. 

\section{Conclusions}
 
We have presented a detailed analysis of the generation of polarization-entangled photons from quantum-dot biexcitons via cavity-assisted two-photon emission. In particular we have studied the dependence of the polarization entanglement on system design and parameters, such as cavity quality and frequency and biexciton binding energy. \textcolor{black}{We have further analyzed photon statistics and spectral properties of the emitted photons.} A focus of our present study lies on the role that different loss mechanisms including pure dephasing, loss into leaky cavity modes, and phonon-assisted cavity feeding at finite temperatures play for the achievable polarization entanglement. \textcolor{black}{Tuning the cavity to half the biexciton energy, for a bound biexciton the biexciton to exciton transition is red-shifted relative to the cavity mode. Therefore at low temperatures with low probability for phonon absorption from the bath, feeding the biexciton-exciton cascade through phonon-assisted processes is strongly suppressed. As a consequence, even in high-quality cavities where phonon-assisted processes are strongest, at low temperature the emission can efficiently be channelled into the two-photon emission process such that a high degree of polarization entanglement is achieved.} Radiative loss reduces the overall quantum efficiency but only slightly alters the entanglement properties of the photons emitted from the system cavity mode. With increasing temperature, a detrimental influence of the coupling of the system to the bath of LA phonons on the achievable polarization entanglement is found.  \\

\begin{acknowledgments}
We acknowledge valuable discussion with Christopher Gies, Matthias Florian, Paul Gartner, and Frank Jahnke from the University of Bremen. We gratefully acknowledge financial support from the DFG through the research center TRR142 and doctoral training center GRK1464, from the BMBF through Q.com 16KIS0114, and a grant for computing time at $\mathrm{PC^2}$ Paderborn Center for Parallel Computing. Stefan Schumacher further acknowledges support through the Heisenberg programme of the DFG.
\end{acknowledgments}

\bibliography{Literatur}

\end{document}